\documentstyle[12pt,aasms4]{article}

\begin{document}

\title{COSMOLOGICAL MAGNETIC FIELD AMPLIFICATION AROUND $Z \sim 100$}

\author{Esteban A. Calzetta and Alejandra Kandus }
\affil{Departamento de F\'\i sica, Fac.Cs. Ex. y Nat. UBA\\
Ciudad Universitaria, (1428) Buenos Aires, Argentina\\
and\\
Instituto de Astronom\'\i a y F\'\i sica del Espacio\\
c.c. 67, Suc. 28 (1428) Buenos Aires, Argentina.}

\begin{abstract}
We study the corrections to the conformal evolution of primordial magnetic
fields after recombination, produced by the fall of the ionized fraction of
matter into the dark matter gravitational wells. This effect enhances the
field around the protostructures being formed, and might help to understand
the fields observed in galaxy clusters and hydrogen clouds.\\
{\it subject headings:} gravitation -  magnetic fields - cosmology: theory.

\end{abstract}

\section{Introduction}

The goal of this paper is to study the evolution of cosmic magnetic fields
interacting with charged matter inmediately after recombination. At this
time most ordinary matter combines in neutral hydrogen, but still a non
negligible ionized fraction remains. On the other hand, as we shall show,
the cosmic microwave background radiation (CMBR) is already too cold to
provide sustantial dissipation, and so the ionized matter, as well as the
neutral one, falls within the gravitational potential wells generated by
dark matter overdensities. The infall of charged matter drags the field
lines, thus resulting in a field amplification in and around the proto
structures being formed. Thus a mechanism results which may provide a
preamplification of a primordial magnetic field, before other best known
processes, such as the galactic dynamo (\cite{ruzmaikin90}), become operational. 
The primeval motivation for the search of these
corrections is to try to understand the presence of magnetic fields of
intensity and coherence similar to the galactic ones, in galaxy clusters and
much less evolved gaseous systems such as the damped Lyman-${\alpha }$
clouds (\cite{oren}). Although the existence and ultimate characteristics of those fields
can hardly be explained by gravitational collapse alone, it is important to
determine if this process may help to relax the requirements on the
mechanisms for primordial field generation.

The existence of relatively intense and coherent magnetic fields in
galaxies, clusters of galaxies and hydrogen clouds is at present theme of
intense research. There is a variety of proposals aimed at explaining their
existence and characteristics, but none is completely successful. They can
be divided in two main sets: one takes the point of view of local
generation, and the other of the amplification of a primordially generated
field (\cite{ruzmaikin90,harrison73,pudritz89,howard96}). While both mechanisms 
compete on equal footing in explaining the
presence of magnetic fields in low redshift objects, like our own galaxies,
the second approach seems more apropriate for highly redshifted, less
developed systems, like the Lyman-$\alpha $ clouds. A model based on
amplification mechanisms requires two stages, one to create the field and
another to amplify it. For the first process there are several proposals
(\cite{dolgov93,dolgov-silk93,ratra92,mazzitelli95,calzetta98,cheng94,sigl97,enqvist94,vachaspati91,hogan83,martin95}), some of which create fields 
strong enough but of
coherence length much smaller than observed, while others generate coherent
fields but of very weak strength. As for the second stage, the ``turbulent
dynamo`` was a preferred amplifying mechanism for a long time 
(\cite{ruzmaikin90}). But once again we must question it after the detection in damped
Lyman-$\alpha $ clouds (\cite{oren,wolfe92}) of magnetic fields of
the same intensity and coherence as the galactic ones. Another promising
mechanism are inverse cascades of magnetic helicity 
(\cite{olesen97,brandenburg97}), but, while this effect is quite well understood in
ordinary magnetohydrodynamics (\cite{moffat61}), its generalization to cosmology
is still an open problem.

It is well known from the theory of Large Scale Structure Formation, that
galaxies are formed due to the infall of barionic matter into the
gravitational potential wells of dark matter, the formation of those last
beginning at the moment of equality between matter and radiation ($z\simeq
20000$). Although they are not ``deep'' wells by the decoupling of barionic
matter and radiation ($z\simeq 1100$), as we know from the anisotropy of the
CMBR, their existence cannot be disregarded. Despite this fact, current
estimates in the literature often assume that, between decoupling and the
starting of the dynamo mechanism or non linear gravitational collapse, their
propagation takes place in a homogeneous, FRW Universe, and
consequently that their evolution law is ${\cal B}\propto B_0/a^2$, where $a$ 
is the scale factor of the Universe and $B_0$ an initial field intensity. In this
work we study in more detail the evolution of a primordial magnetic field in
a matter dominated, inhomogeneous Universe, with the aim of estimating the
corrections to this conformal evolution due to the evolution of the
background perturbations.

Long time ago Ya. B. Zel'dovich et al (1983) investigated the effects of a
magnetic field over anisotropic self gravitational collapse of matter. 
They also estimated the amplification that a
primordial field would undergo by that collapse, finding a factor of ${\cal O
}\left( 10^3-10^4\right) $. Working backwards from present estimates, the
inferred intensity for a primordial field results then $B_0\simeq
10^{-9}-10^{-10}$ Gauss, much larger than the ones predicted by the
available mechanisms of primordial generation, for the required coherence
scale.

Our proposal differs from Zel'dovich's in that we do not study self
gravitational collapse, but rather the infall of charged matter into
preexistent potential wells created by dark matter. Consider a spatially
flat Friedmann - Robertson - Walker Universe endowed with adiabatic
perturbations. The main matter component is cold dark matter, which for our
purposes only means that the smallest structures to be formed are galaxies.
At the moment of equilibrium between matter and radiation, the matter
dominated era of the Universe begins and the adiabatic perturbations in the
dark matter component can grow, creating an inhomogeneous Universe. On the
other side, the rest of the matter component, the baryons, remain coupled to
photons until the temperature of the Universe is low enough to let electrons
and protons combine. Then neutral hydrogen emerges while a small ionized
fraction of ${\cal O}\left( 10^{-5}-10^{-7}\right) $ remains.

Any magnetic field present in the Universe will evolve coupled to the
charged matter plasma. This means that before recombination, the field will
be tied to the electron-photon plasma, and will remain coupled to the
remaining ionized matter after that moment. Before recombination the plasma
is hot enough to be considered relativistic and therefore its evolution can
be followed without taking into consideration the background
inhomogeneities. After recombination, however, the situation is the reverse:
the temperature is too low to neglect the effect of the perturbations and
therefore we must study the evolution of the magnetic field coupled to the
remaining ionized matter, in an inhomogeneous Universe. Of course the
ionized plasma will be coupled to the CMBR as well as to any primordial
field; one of our objectives is to show that, in the era under
consideration, the interaction with the CMBR does not affect sustantially
the collapse process, and that our plasma can be consider as a perfect fluid
with infinite conductivity.

Several authors have studied related physical situations. For example
Ryu et al (1998) and Blasi et al (1999) studied the
effect of inhomogeneities in the matter distribution of the Universe on the
Faraday rotation of radio signals and light from quasars respectively. The
inhomogeneities considered by the former authors are large scale filaments
and sheets (generated by a hydrodynamical code), to which the magnetic field 
is glued. They assumed that a seed field is locally generated by a battery 
mechanism and further amplified by streaming and turbulent motions.
Blasi et al, on the other side, considered redshifts for which the
inhomogeneities are well described by the Ly-$\alpha $ forest. The magnetic 
field is also glued to this structure and they assume it scales with the electron 
density as $B \propto n_2^{2/3} $. In both works an upper limit for the magnetic 
field strength in the structures is derived.

Our work differs substantially from the one by Blasi et al (1999) in that we 
do not assume a given relationship between magnetic field
and ionized matter density, rather we derive it. 
The difference with the work by Ryu et al (1998) is that we do not
assume a locally generated magnetic field, but a primordially generated one.
Besides, we study the evolution of the magnetic field in a universe permeated by
primordial inhomogeneities that are still in their linear regime of evolution, 
which means higher redshifts than in the mentioned works.

We have therefore the following scenario: a weak primordial magnetic field
is coupled to the matter that remains ionized after the recombination of
hydrogen, and the whole system propagates in an inhomogeneous universe. The
matter component falls into the background overdensities due to
gravitational attraction, dragging with it the magnetic field lines. This
drag produces a local enhancement of the magnetic field in the overdense
regions, deviating from the evolution to be expected in a pure FRW Universe.
Our purpose is to calculate this enhancement and to see if it is intense
enough to explain (up to certain level) the detected field intensities.

The paper is organized as follows: In section {\bf 2} we describe the
geometry of the background Universe, in section {\bf 3} we describe the
matter component the magnetic field is tied to, discussing in some detail
the possible interactions with the CMBR. In section {\bf 4} we study the
evolution of the magnetic field and summarize our main conclusions in
Section {\bf 5}.

\section{Baryon free fall}

\subsection{The geometry of the background}

Consider a matter dominated flat FRW Universe, endowed with adiabatic
perturbations. In the longitudinal gauge, the metric tensor for this
Universe reads (\cite{branden92}) 
\begin{equation}
g_{\mu \nu }=a^2\left( \eta \right) \left( 
\begin{array}{cc}
-\left( 1+2\psi \right) & 0 \\ 
0 & \left( 1-2\psi \right) \gamma _{ij}
\end{array}
\right)  \label{aa}
\end{equation}
where $a\left( \eta \right) =(\eta /\eta _0)^2$ is the scale factor, $\eta $
the conformal time and $\gamma _{ij}$ the spatial flat three metric, $\gamma
_{ij}=diag(1,1,1)$. We have normalized the conformal factor to unity at the
time of decoupling, given by $\eta =\eta _0\simeq 0.4h^{-1}\times
10^{38}GeV^{-1}.$ Greek indices run from $0$ to $3$, while latin ones take
the values $1,2,3$. The function $\psi $ depends on $r$ and $\eta $ and can
be interpreted as a generalized newtonian potential. Its amplitude can be
determined by the level of anisotropies in the background radiation caused
by the Sachs-Wolfe effect at decoupling. If $\psi $ describes a growing
perturbation then it must depend only on $r$ (\cite{branden92}). In
the linear regime, this mode will grow as $\sim a(\eta )\psi (r)$.

From this expression we can find the four velocity of the background fluid
flow and construct a $3+1$ description. This is needed because we will
consider the presence of a magnetic field, and its definition relays on the
specification of the spatial surfaces of the spacetime.

In a coordinate system where peculiar velocities are zero we have

\begin{equation}
U^\mu =\frac{dx^\mu }{ds}=\left( \frac{1-\psi }a,\vec 0\right)  \label{ab}
\end{equation}
This is normalized by 
\begin{equation}
U^\mu U_\mu =-1  \label{ac}
\end{equation}
Finally we build a projection tensor for the spatial three surfaces as 
\begin{equation}
h^{\mu \nu }=g^{\mu \nu }+U^\mu U^\nu  \label{ad}
\end{equation}

As this fluid has zero vorticity, the spatial three surfaces are orthogonal
to the fluid four velocity (\cite{vanelst98})

\subsection{The plasma component of the Universe}

We want to study the evolution in the matter dominated era of the Universe
of a primordially generated magnetic field, considering that in that period
of time the process of structure formation is taking place as well. At the
beginning of matter dominance ($z\simeq 20000$), the background
perturbations begin to grow, while the baryons remain coupled to the cosmic
background radiation. When the temperature of the Universe is low enough,
electrons and protons recombine ($z\simeq 1100$) and most of the barionic
matter becomes electrically neutral. It can then start falling into the
already formed dark matter overdensities, which are described (in the linear
regime) by the function $\psi (r)$. There remains a ionized fraction of
barionic matter of order $\sim {\cal O}(10^{-5}-10^{-7})$.

We assume that any magnetic field present in the Universe interacts only
with the barionic component of the matter. With the electrically neutral
fraction, the interaction is very weak and is described by ${\bf B.\bf m}$
where $\vec m$ is the atomic magnetic moment. With the ionized fraction, the
interaction is given by the Lorentz force, being therefore stronger than the
former. We will study the system of ionized barionic matter
coupled to a preexisting magnetic field, in the matter dominated epoch, by
means of the equation 
\begin{equation}
{\cal T}_{\;\;;\nu }^{\mu \nu }={\cal F}^\mu  \label{ba}
\end{equation}
where ${\cal T}^{\mu \nu }$ is the stress energy tensor for the matter and $
{\cal F}^\mu $ is the Lorentz force density, exerted by the magnetic field
on the charged particles.

We shall investigate the effects on the preexisting magnetic field of the
free fall of matter into the newtonian potential wells described by $\psi
\left( x\right) .$ To this end, we shall consider a geometry where $\psi
=\psi \left( z\right) $, and correspondingly the motion of matter is along
the $z$ direction only. In this first Section, moreover, we shall make two
simplifying assumptions, namely, we shall consider the plasma as a perfect
fluid, and we shall neglect the right hand side of Eq. (\ref{ba}). We shall
show the validity of the first approximation towards the end of this
Section; the second one will put limits on how far the model can be relied
upon as a description of collapse, and will be discussed later on.

We shall treat the energy density $\delta \varepsilon $, pressure $\delta p$
and four velocity $\delta u^\mu $ of the ionized plasma as perturbations on the
corresponding quantities for the dark matter, namely $\epsilon =\varepsilon
+\delta \varepsilon $, $P=p+\delta p,$ and ${\cal U}^\mu =U^\mu +\delta
u^\mu $. We shall call $\delta {\cal N}$ any conserved quantity proper of the ionized
fraction, for example the leptonic or barionic numbers, and we will consider that
this quantity does not vary with time, i.e. the ionisation fraction remains
constant (\cite{luchin,novikov83})
Since the gravitational effect of this perturbation is negligible,
we shall not consider the perturbations to the geometry that it could produce.
We are therefore describing the free fall of the plasma into the
potential wells created by the dark matter. The full energy - momentum
tensor for the plasma is 
\begin{equation}
{\cal T}^{\mu \nu }=T_I^{\mu \nu }+\delta T_I^{\mu \nu }  \label{bb}
\end{equation}
where
\begin{equation}
T_I^{\mu \nu }=\varepsilon U^\mu U^\nu +ph^{\mu \nu }  \label{bc}
\end{equation}
\begin{equation}
\delta T_I^{\mu \nu }=\delta \varepsilon U^\mu U^\nu +\left( \varepsilon
+p\right) \left( \delta u^\mu U^\nu +U^\mu \delta u^\nu \right) +\delta
ph^{\mu \nu }  \label{bd}
\end{equation}

As stated before, we have 
\begin{equation}
{\cal T}_{;\nu }^{\mu \nu }=0  \label{be}
\end{equation}

We will also need the equation for the conservation of the number of 
matter particles, namely
\begin{equation}
{\cal I}^{\mu }_{;\mu } = 0 \label{bf}
\end{equation}
where
\begin{equation}
{\cal I}^{\mu } =  \delta {\cal N} \left( U^{\mu } + 
\delta u^{\mu }\right) \label{bg}
\end{equation}

\subsection{Timelike and spacelike projections of ${\cal T}^{\mu \nu}_{\; ;\nu }$}

Our ultimate goal is to study the evolution of the matter component and the
magnetic field under the gravitational perturbations created by dark matter.
As the specification of what magnetic and electric fields are is tied to the
definition of spacelike surfaces and their corresponding timelike curves, we
must work in a $3+1$ formalism. For this purpose we project equation (\ref{be}) 
onto the spatial surfaces, by $h_\nu ^\mu $ (eq.(\ref{ad})) and along
the worldlines given by $U^\mu $ (eq. (\ref{ab})). Performing the covariant
derivatives with the Christoffel symbols corresponding to the metric tensor
Eq. (\ref{aa}) (see Appendix) and replacing the explicit expressions for the
four velocity we obtain the conservation laws to leading and first order as
follows:

$zeroth\;order$

\begin{equation}
\dot \varepsilon +3\frac{\dot a}a\left( \varepsilon +p\right) =0  \label{ca}
\end{equation}

For the dark matter component, moreover, $p=0$, and we get the expected
behavior $\varepsilon \sim a^{-3}.$

$first\;order$
\begin{equation}
\ \ \frac \psi a\dot \varepsilon -\frac 1a\delta \dot \varepsilon +3\left[ 
\frac{\dot a}{a^2}\psi +\frac{\dot \psi }a\right] \left( \varepsilon
+p\right) -3\frac{\dot a}{a^2}\left( \delta \varepsilon +\delta p\right)
-\left( \varepsilon +p\right) \delta u^j,_j=0  \label{cb}
\end{equation}
\begin{equation}
\ \frac{\psi ^{,i}}{a^2}\left( \varepsilon +p\right) +\frac{\delta p^{,i}}{
a^2}+\left( \dot \varepsilon +\dot p\right) \frac{\delta u^i}a+\left(
\varepsilon +p\right) \left\{ \frac 1a\delta \dot u^i+5\frac{\dot a}{a^2}
\delta u^i\right\} \ =0  \label{cc}
\end{equation}

We now replace the zeroth order equation and write $\delta \varepsilon =\rho
\varepsilon $, $\delta u^j=u^j/a$ and $p=c_s^2\varepsilon $, where $c_s$ is
the speed of the sound in the medium. We will also neglect the terms with $
\dot \psi $ as they correspond to decaying modes of the background
perturbations, leading to 
\begin{equation}
\dot \rho +\left( 1+c_s^2\right) u^j,_j=0  \label{cd}
\end{equation}
\begin{equation}
\ \ \ \psi ^{,i}+\frac{c_s^2}{\left( 1+c_s^2\right) }\rho ^{,i}+\left(
1-3c_s^2\right) \frac{\dot a}au^i+\dot u^i=0  \label{ce}
\end{equation}

In what follows, we neglect the effect of the pressure. This is justified in
view of the low temperature of the matter.

\subsection{Particle number conservation}

Developing the covariant derivative of eq. (\ref{bf}) we obtain
\begin{equation}
\delta \dot {\cal N} \left( 1 - \psi \right) +
3 \delta {\cal N} \frac {\dot a}a \left( 1 - \psi \right)
+ \left( \delta {\cal N} u^j\right) ,_j - 
2 \delta {\cal N} \psi ,_j u^j = 0
\label{cf}
\end{equation}

 Taking 
$ \delta {\cal N} =  \delta n /a^3$ the previous equation reads
\begin{equation}
\left( 1 - \psi \right) \delta \dot n + \left[ u^j \delta n\right] ,_j 
-2\delta n \psi ,_j u^j= 0 \label{ch}
\end{equation}

\subsection{Study of the evolution of $\rho $ and $u^j$ for a dustlike gas}

We are now ready to analyze the effects that the background perturbations
may have on the density and velocity of the plasma. Current numerical
investigations of large scale structure formation show that gravitational
colapse of matter is highly anisotropic, giving rise to a 
complicated network of rich and poor clusters connected by 
filaments with void regions in between (\cite{bond96}). 
This is a difficult process to study analytically so we will retain one simple
feature, namely that gravitational collapse is not symmetric, but takes place 
mainly along a preferred direction, the resulting structures being pancake or 
cigar-like shaped. This fact allows us to introduce another simplifying 
hypothesis: we will consider that all quantities depend only on the coordinate 
along which gravity acts, let it be the $z$ coordinate. 
This simplification contains a very important physical assumption about the 
perturbations: they are vorticity free. Were we to consider two or three 
dimensional gravitational collapse, vorticity must be taken into account.

When we make the mentioned approximations we are left with 
\begin{equation}
\dot \rho +u^j,_j=0  \label{da}
\end{equation}
\begin{equation}
\psi ^{,i}+\frac{\dot a}au^i+\dot u^i=0  \label{db}
\end{equation}
We write ${\bf u}=\left( u_x,u_y,u_z\right) $and obtain 
\begin{equation}
\dot \rho +\partial _zu_z=0  \label{dc}
\end{equation}
\begin{equation}
\dot u_x+\frac{\dot a}au_x=\dot u_y+\frac{\dot a}au_y=0  \label{dd}
\end{equation}
\begin{equation}
\partial _z\psi +\dot u_z+\frac{\dot a}au_z=0  \label{de}
\end{equation}

For the initial conditions, ${\bf u}(z,\eta _0)=0$, and $\rho (z,\eta
_0)=\rho _0$, the solutions to equations (\ref{dc}) and (\ref{de}) read 
\begin{equation}
\rho =\rho _0+\frac{\eta _0^2}3\partial _z^2\psi (z)\left[ \frac 12\left( 
\frac{\eta }{\eta _0} \right) ^2+\frac {\eta _0}{\eta }-\frac 32\right]
\label{df}
\end{equation}

\begin{equation}
u_z(z,\eta )=\frac{\eta _0}3\partial _z\psi (z)\left[ \left( \frac{\eta _0}
\eta \right) ^2-\frac \eta {\eta _0}\right]  \label{dg}
\end{equation}
\begin{equation}
u_x=u_y=0  \label{dh}
\end{equation}
These are the equations which describe the free fall of the plasma into the
potential wells.

\subsection{Discussion}

Equations (\ref{ch}) and (\ref{db}) are the goal of this Section, namely,
they describe the evolution of the number density
and velocity of the plasma as it falls on the dark matter overdensities. 
Since the magnetic field lines are tied to the charges, we expect these 
will be dragged along with matter, producing an amplification of the 
field in the overdense regions. The study of this effect is the subject 
of the rest of this paper.

However, before we go on it is right that we discuss the first of the
approximations under which eqs. (\ref{da}) and (\ref{db}) were derived,
namely, the assumption that the plasma could be described as a perfect
fluid. In reality, the plasma interacts not only with the gravitational
field of the dark matter, but also with the photons in the CMBR, and this
interaction could in principle result in dissipative behavior.

In the epoch we are interested in, the plasma and the CMBR are not in
equilibrium, and therefore the usual estimates of the dissipative effects
induced on the plasma do not apply (\cite{weinberg71}). Indeed, the mean free
path for the photons is $l_{mfp}=\left( n_c\sigma _T\right) ^{-1}$, where $
n_c$ is the number density of the electrons in the ionized fraction of
matter and $\sigma _T$ $=8\pi e^2/3m_c^2$ is the Thompson cross section,
(due to the low temperature of the background photons, the main interactive
process is Thompson scattering). Their values are $\sigma _T\simeq 1.2\times
10^5GeV^{-2}$, $n_c\simeq 0.4\times 10^{-45}a(\eta )^{-3}GeV^3$ and
therefore $l_{mfp}\simeq 2\times 10^{40}a(\eta )^3GeV^{-1}$. On the other
side, the size of the particle horizon is $d_h=\eta ^3/\eta _0^2$. We
estimate the number of collisions that a photon suffers per Hubble time as $
n_{coll}=d_h/l_{mfp}\sim 20\,a(\eta )^{-3/2}$, which is too low to sustain
equilibrium.

Still, the plasma must feel some kind of friction as it moves accross the
CMBR. Since the plasma is electrically neutral, any baryon motion is matched
by a corresponding electron flux. The Thompson cross section is $10^{-6}$
orders of magnitude smaller for protons than for electrons, then the
interactions of photons with electrons dominate, and we can consider that
the electrons suffer a friction due to the photons. Therefore, instead of
the conservation equation (\ref{be}) we should write

\begin{equation}
{\cal T}_{\;;\nu }^{\mu \nu }=\zeta ^\mu ;\;{\cal I}_{;\mu }^\mu =0  \label{ea}
\end{equation}
where ${\cal I}^\mu $ is the particle four current. We also have the corresponding
equation for the photons 
\begin{equation}
T_{\gamma ;\nu }^{\mu \nu }=-\zeta ^\mu  \label{eb}
\end{equation}

From the first principle of thermodynamics we have 
\begin{equation}
S^\mu =\Phi ^\mu -\beta _{c\nu }{\cal T}^{\mu \nu }-\beta _{\gamma \nu
}T_\gamma ^{\mu \nu }-\alpha {\cal I}^\mu  \label{ec}
\end{equation}
where $S^\mu $ is the entropy flux, $\beta _c^\mu ={\cal U}^\mu /T_c$, $
\beta _\gamma ^\mu =U^\mu /T_\gamma $ , the subindex $c$ corresponding to
the ``charged'' matter and $\gamma $ to the CMBR. $\Phi ^\mu $ is the
thermodynamic potential (Israel 1988), whose derivatives are 
\begin{equation}
\frac{\partial \Phi ^\mu }{\partial \beta _{c\nu }}={\cal T}^{\mu \nu };\;
\frac{\partial \Phi ^\mu }{\partial \beta _{\gamma \nu }}=T_\gamma ^{\mu \nu
};\;\frac{\partial \Phi ^\mu }{\partial \alpha }={\cal I}^\mu .  \label{ed}
\end{equation}

By the second law of thermodynamics we have 
\begin{equation}
S_{;\mu }^\mu =\left( \beta _{\gamma \nu }-\beta _{c\nu }\right) \zeta ^\nu
>0  \label{ee}
\end{equation}
which is satisfied only if 
\begin{equation}
\zeta ^\mu =C_1\beta _\gamma ^\mu \left( -\beta _\gamma ^2-\beta _{\gamma
\mu }\beta _c^\mu \right) +C_2\left( \beta _\gamma ^\nu -\beta _c^\nu
\right) \left[ \delta _\nu ^\mu +\frac {\beta _\gamma ^\mu
\beta _{\gamma \nu }}{\beta _\gamma ^2}\right]  \label{ef}
\end{equation}
where $\beta_{\gamma }^2 = -\beta_{\gamma \mu} \beta _{\gamma }^{\mu }$ and 
$C_1,C_2 \geq 0$. 
In terms of the four velocities and temperatures eq. (\ref{ef}) reads 
\begin{equation}
\zeta ^\mu =-\frac{C_1U^\mu }{T_\gamma T_c}\left[ T_c+T_\gamma U_\nu {\cal U}
^\nu \right] -\frac{C_2}{T_c}\left[ {\cal U}^\mu +U^\mu U^\nu {\cal U}_\nu
\right]  \label{eg}
\end{equation}
We see that the first term in the r.h.s. of eq. (\ref{eg}) can be
interpreted as the heat interchange between the photons and the plasma,
while the second one as the momentum transfer between them.

To estimate the $C_1$ and $C_2$ coefficients, let us place ourselves in the
rest frame of the CMBR. Since in this frame the photon bath is isotropic,
the bombardment of electrons ''at rest'' by photons averages out, and any
net effect is solely due to the motion of the electrons. Each time an
electron strikes a photon, the later gains a momentum $\Delta p_\gamma \sim
1/\lambda \sim T_\gamma $. The electron loses a momentum $\Delta p_e\sim
-T_\gamma $ and changes its energy by $\Delta U=-v_eT_\gamma $, where $v_e$
is the velocity of the electrons. The number of collissions per unit time
that an electron suffers is $dn=n_\gamma \sigma _Tv_e$. If the electron
number density is given by $n_e$, then the net force density excerted by the
photons on the electrons is given by $F_{coll}^e\simeq n_en_\gamma \sigma
_TT_\gamma v_e$, and the mean energy loss is $Q=n_en_\gamma \sigma
_TT_\gamma v_e^2$. Performing the average over all electrons, and comparing
with eq. (\ref{eg}), we conclude 
\begin{eqnarray}
\frac{C_1}{T_\gamma T_c} &=&\frac{n_cn_\gamma \sigma _TT_\gamma }{m_e}
\label{eh} \\
\frac{C_2}{T_c} &=&n_cn_\gamma \sigma _TT_\gamma  \label{ei}
\end{eqnarray}

We should now project eq. (\ref{eg}) along the fluid flow lines and onto the
orthogonal spatial surfaces and replace the expressions in the r.h.s. of
eqs. (\ref{cb}) and (\ref{cc}) respectively. For the timelike projection
only the first term of eq. (\ref{eg}) contributes while for the spacelike
projection we need the second term of that equation. Considering $
\varepsilon $ as the critical density of the Universe, using also that $
\varepsilon _{crit}=\varepsilon _{crit}^{(0)}/a^3(\eta )$, $
n=n^{(0)}/a^3(\eta )$ and that $T=T^{(0)}/a(\eta )$, where with the
supraindex $(0)$ we refer to quantities at the epoch of recombination, we
have that the equations (\ref{da}) and (\ref{db}) now read

\begin{equation}
\dot \rho +u^j,_j=-\frac{n_c^{(0)}n_\gamma ^{(0)}\sigma _TT_\gamma ^{(0)}}{
m_e\varepsilon _{crit}^{(0)}a^3(\eta )}\left( T_c-T_\gamma \right)
\label{ej}
\end{equation}
\begin{equation}
\psi ^{,i}+\frac{\dot a}au^i+\dot u^i=-\frac{n_c^{(0)}n_\gamma ^{(0)}\sigma
_TT_\gamma ^{(0)}}{\varepsilon _{crit}^{(0)}a^5(\eta )}u^i  \label{ek}
\end{equation}
Replacing the figures, $\varepsilon _{crit}^{(0)}\sim 10^{-38}$ GeV$^4$ $
n_\gamma ^{(0)}\sim 10^{-33}$ GeV$^3$, $\sigma _T\sim 10^5$ GeV$^{-2}$, $
T_\gamma ^{(0)}\sim 10^{-10}$ GeV, we obtain for the r.h.s. of eq. (\ref{ej}
) 
\begin{equation}
\frac{n_\gamma ^{(0)}n_c^{(0)}\sigma _TT_\gamma ^{(0)2}}{m_e\varepsilon
_{crit}^{(0)}a^4(\eta )}\simeq \frac{10^{-49}}{a^3(\eta )}GeV  \label{el}
\end{equation}
where we have assumed $T_c\ll T_\gamma $. We compare this quantity with the
divergence of the velocity in the l.h.s. of eq. (\ref{da}), using eq. (\ref
{dg}). Asuming $\partial _z\sim z_0^{-1}$ where $z_0$ is a characteristic
scale of the background inhomogeneity, which for a galaxy is $z_0^G\sim
10^{35}$ GeV$^{-1}$, we have $\eta _0/z_0^G\sim 10^2$. Recalling that
initially the peculiar velocity is zero, we have that the effect of the heat
transfer becomes negligible inmediately after the matter starts falling into
the dark matter overdensities.

For equation (\ref{db}) we compare the second term in the l.h.s. with the
r.h.s.. For the first we have that the coefficient of $u^i$ reads $\dot
a(\eta )/a(\eta )=2/\eta =2\eta _0^{-1}a^{-1/2}(\eta )$. For the coefficient
of $u^i$ in the r.h.s. we have 
\begin{equation}
\frac{n_c^{(0)}n_\gamma ^{(0)}\sigma _TT_\gamma ^{(0)}}{\varepsilon
_{crit}^{(0)}a^5(\eta )}\simeq \frac{10^{-45}}{a^5(\eta )}GeV  \label{fd}
\end{equation}

This factor will be smaller than the corresponding one in the l.h.s. of eq. (
\ref{db}) when $a(\eta )^{9/2}\geq 1$, and this constraint is always
satisfied, because we have $a(\eta )\geq 1$. We therefore have that in the
whole time interval considered in the paper, the equations for a perfect
fluid apply.

\section{Evolution equation for the magnetic field}

As we have seen in the previous Section, after decoupling the partially
ionized plasma falls into the newtonian potential wells created by dark
matter. In its infall, it will dragg the magnetic field lines, producing an
enhancement of the field intensity, proportional to the depth of the
potential well. Our goal in this Section is to study this effect. In the
last subsection, we will discuss the limitations on our analysis.

\subsection{Maxwell Equations}

Let us begin by deriving the Maxwell Equations that satisfy the
electromagnetic field in the spacetime we are working in. These were put in
the $3+1$ form for the first time by G. F. R. Ellis (1973) and latter by K.
Thorne \& D. MacDonald (1982).

We begin by defining the field strength tensor $F_{\mu \nu }$ as usual 
\begin{equation}
F_{\mu \nu }=\partial _\mu A_\nu -\partial _\nu A_\mu  \label{ga}
\end{equation}
where $A^m$ is the electromagnetic four potential. The electric and magnetic
fields are given by 
\begin{eqnarray}
{\cal E}_\mu &=&F_{\mu \nu }U^\nu =-F_{\nu \mu }U^\nu  \label{gb} \\
{\cal B}_\mu &=&\frac 12\eta _{\mu \nu \alpha \beta }U^\nu F^{\alpha \beta }
\label{gc}
\end{eqnarray}
We can rewrite the field strength tensor in terms of these fields as 
\begin{equation}
F_{\mu \nu }=U_\mu {\cal E}_\nu -{\cal E}_\mu U_\nu -\eta _{\mu \nu \alpha
\beta }U^\alpha {\cal B}^\beta  \label{gd}
\end{equation}

and with the indices up 
\begin{equation}
F^{\mu \nu }=U^\mu {\cal E}^\nu -{\cal E}^\mu U^\nu -\eta _{\;\;\;\alpha
\beta }^{\mu \nu }U^\alpha {\cal B}^\beta  \label{ge}
\end{equation}
where $U^\mu $ is the four velocity of fiducial observers, which in our case
is given by equation (\ref{ab}). The Maxwell equations can be written in
covariant form as 
\begin{eqnarray}
F_{\;\;\;;\nu }^{\mu \nu } &=&{\cal J}^\mu  \label{gf} \\
\eta ^{\mu \nu \alpha \beta }F_{\nu \alpha ;\beta } &=&0  \label{gh}
\end{eqnarray}
We will need the projections of these equations onto spatial surfaces and
along the four velocity. We also transform the electromagnetic field and
current as ${\cal E}^i\rightarrow E^i/a^2$, ${\cal B}^i\rightarrow B^i/a^2$, 
${\cal J}^i\rightarrow J^i/a^3$. The previous set of equations then reads

\begin{equation}
-E^k,_k+3\psi ,_kE^k=-\left( 1+\psi \right) J^0  \label{gi}
\end{equation}

\begin{equation}
-\dot E^i\left( 1-\psi \right) +\eta _{\;\;\;k0}^{ij}\left[ \left( 1-\psi
\right) B^k,_j-3\psi ,_jB^k\right] =J^i  \label{gj}
\end{equation}

\begin{equation}
B_{\;;j}^j-B^j\psi ,_j=0  \label{gk}
\end{equation}

\begin{equation}
\eta _{\;\;\;k0}^{ij}\left[ \left( 1-\psi \right) E^k,_j-3\psi ,_jE^k\right]
+\left( 1-\psi \right) \dot B^i=0  \label{gl}
\end{equation}
These are the equations that we will use in the main part of this work.

\subsection{Plasma conductivity and the magnetic field}

The currents and fields appearing in Maxwells equations above are further
related by Ohm's law 
\begin{equation}
{\cal J}^\mu +{\cal U}^\mu {\cal U}^\nu {\cal J}_\nu =\sigma F^{\mu \nu }
{\cal U}_\nu  \label{ha}
\end{equation}
where ${\cal U}^\mu $ is the fluid four velocity given by ${\cal U}^\mu
=U^\mu +\delta u^\mu $. Keeping only first order terms we have 
\begin{equation}
J^i-u^i\rho _e=\sigma \left[ E^i+\eta _{\;\;pr}^{i0}u^pB^r\right]  \label{hb}
\end{equation}

In our case, it is appropriate to consider the plasma as perfectly
conducting. This is essentially due to the fact that any charge separation
would lead to an electrostatic attraction much stronger than the
gravitational forces considered so far, and thus it cannot be sustained over
any macroscopic lapse. For an infinite conductivity, the electric field in
the rest frame of the plasma must vanish, and this yields an extra relation
among the electric and magnetic fields, and the plasma velocity in that frame, 
namely 
\begin{equation}
E^i=-\eta _{\;\;pr}^{i0}u^pB^r  \label{hc}
\end{equation}
For equation (\ref{gl}) we have 
\begin{equation}
\left( 1-\psi \right) {\bf \nabla } \times \left( {\bf u}\times {\bf B}\right)
-3{\bf \nabla } \psi \times \left( {\bf u}\times {\bf B}\right) =\left( 1-\psi
\right) \frac{\partial {\bf B}}{\partial \eta }  \label{hd}
\end{equation}
This is the evolution equation for the magnetic field.

Besides the arguments above, it is possible to give a direct estimate of the
plasma conductivity that confirms it may be considered as infinite. Suppose
an electric field were imposed on the plasma, generating an electron flux
with velocity $v_e$ and a current $j=en_ev_e$. The field yields a Joule
power $P=Ej$, which must be equal to the power dissipated into the CMB by
the friction force (recall the discussion in the previous Section) 
\begin{equation}
P=2n_en_\gamma \sigma _Tv_e^2T=\frac{2n_e^0n_\gamma ^0\sigma _Tv_e^2T^0}{
a\left( \eta \right) ^9}  \label{he}
\end{equation}
Therefore 
\begin{equation}
v_e=\frac{eE}{Tn_\gamma \sigma _T}  \label{hf}
\end{equation}
from where we read the conductivity 
\begin{equation}
\sigma _c=\frac{e^2n_e}{Tn_\gamma \sigma _T}  \label{hg}
\end{equation}
Estimating its value with the same figures as before we obtain 
\begin{equation}
\sigma _c\sim 10^{-11}a\left( \eta \right) GeV  \label{hh}
\end{equation}
In order to determine if this conductivity is large enough to be considered
as infinite, we may compare the conduction current with the displacement
current generated by the collapse, $\partial _\eta E\sim E/\eta _0$, where $
\eta _0^{-1}\sim 10^{-38}GeV.$ We see that we can consider the conductivity
of the plasma as infinite for all practical purposes.

\section{Evaluation of the magnetic field}

The evolution equation for the magnetic field can be simplified as follows:
Write ${\bf u}=\left( 0,0,u_z\right) $ and ${\bf B}={\bf B}\left( \eta
,z\right) $. Then 
\begin{equation}
{\bf \nabla } \times {\bf B}=\left( -\partial _zB_y,\partial _zB_x,0\right)
\label{hi}
\end{equation}
\begin{equation}
{\bf u}\times {\bf B}=\left( -u_zB_y,u_zB_x,0\right)  \label{hj}
\end{equation}
\begin{equation}
{\bf \nabla }\times \left( {\bf u}\times {\bf B}\right) =\left( -\partial
_z\left[ u_zB_x\right] ,-\partial _z\left[ u_zB_y\right] ,0\right)
\label{hk}
\end{equation}
\begin{equation}
{\bf \nabla }\psi \times \left( {\bf u}\times {\bf B}\right) =
\left( -\psi ,_z u_zB_x ,-\psi ,_zu_zB_y ,0\right) \label{hkk}
\end{equation}
We see that $B_z$ is not affected by the collapse. For $B_x$ and $B_y$
we can rewrite equation (\ref{hd}) as (ommiting the subindex)
\begin{equation}
\partial _{\eta }\left[ \left( 1 - \psi \right) ^3 B \right] +
\partial _z\left[ \left( 1 - \psi \right) ^3 B u^z\right] = 0
\label{hl}
\end{equation}
which expresses the conservation of the quantity 
$\left( 1 - \psi \right) ^3B$. 

At this point it is important to note that we cannot
recast equation (\ref{ch}), which expresses the conservation of the number
density that characterizes the ionized fraction, in a form similar to 
eq. (\ref{hl}). This means that the relation between $\delta n$ and $B$
might not be a simple one. Nevertheless since $\psi $ is a small deviation
from inhomogeneity, we can obtain a conservation law from
eq. (\ref{ch}) if we neglect the last term and the newtonian potential 
in the first term. We obtain
\begin{equation}
\delta \dot n + \left( u^z \delta n\right) ,_z = 0 \label{hll}
\end{equation}
where 
\begin{equation}
u^z=f(z)\left[ \left( \frac{\eta _0}\eta \right) ^2-\frac \eta {\eta _0}\right]
\label{hm}
\end{equation}
with $f(z)=\eta _0\partial _z\psi (z)/3$. Performing the same approximations in
equation (\ref{hl}), this equation reads
\begin{equation}
\dot B + \left[ B u^z\right] ,_z = 0
\label{hn}
\end{equation}
Since both $B$ and $\delta n$ are assumed to be homogeneous at $\eta _0$,
from eqs. (\ref{hll}) and (\ref{hn}) we may conclude that $B \propto \delta n$,
at all times, where the constant of proportionality is given by the value of 
$B/\delta n$ at $\eta = \eta _0$, i.e. we have
\begin{equation}
B\left( \eta ,z\right) = \frac {B_0}{\delta n_0}\delta n \left( \eta ,z\right)
\label{ho}
\end{equation}
Observe that this is not the behaviour expected for a homogeneous universe, where
$u^j = 0$ and consequently the solution to eq. (\ref{hn}) would be $B = const$.
The solution to eq. (\ref{hll}) reads
\begin{equation}
\delta n = \frac 1{f(z)} H\left[ \int ^z \frac{dz'}{f(z')} +\eta _0
           \left( \frac 12\frac{\eta ^2}{\eta _0^2} + \frac {\eta _0}{\eta }
           - \frac 32 \right) \right] \label{hp}
\end{equation}
where $H\left( \zeta \right) $is an arbitrary function obtained from the value
of $\delta n$ at $\eta = \eta _0$. The hypothesis that the magnetic field 
$B_0$ is initially uniform over the horizon size, although simplistic, is valid 
for weak magnetic fields (\cite{calzetta98}) and also helps to simplify the 
calculations. The real initial large scale structure of magnetic fields is 
still under study. Jedamzik et al (1998) have shown that linear magnetohydrodynamic 
modes suffer damping during recombination and neutrino decoupling, while
Subramaninan and Barrow (1998) analyzed the non-linear case. They both
find scales below which those modes are dissipated. Nevertheless their
treatment refers to perturbations to a background magnetic field, while here
we aim to study the evolution of that background field in an
inhomogeneous universe.

In order to achieve our purpose, we only need the newtonian  potential 
profile $\psi (z)$ and 
to our ends, it is enough to consider that the dark matter is still in the
linear regime (see below), the shape of the potential being then
essentially given by the theory of primordial density generation 
(\cite{bardeen83,guth82,starob82,hawking82}). However, this theory does 
not yield a deterministic prediction, but only the relative probabilities 
of different density contrast profiles. Therefore we shall restrict 
ourselves to considering a few simple situations, namely, a power law 
density profile, a uniform slab of finite height, and a harmonic density 
contrast. 

\paragraph{Power law density contrast}

Let us consider dark matter density profiles of the form 
\begin{equation}
\delta (z)=\left\{ 
\begin{array}{ccc}
\xi \left( \frac z{z_0}\right) ^{p-2} & for & z<z_0 \\ 
0 & for & z>z_0
\end{array}
\right.  \label{ia}
\end{equation}

They give rise to power law (generalized) Newtonian potentials, i.e.: 
\begin{equation}
\psi (z)=\left\{ 
\begin{array}{ccc}
\alpha \left( \frac z{z_0}\right) ^p & for & z<z_0 \\ 
\alpha \left( p\frac z{z_0}-p+1\right) & for & z>z_0
\end{array}
\right.  \label{ib}
\end{equation}
where $\alpha $ has dimensions of $length^{-1}$ and its amplitude is $\alpha
\sim 10^{-5}$,.and $z_0$ determines the spatial extension of the
perturbation, which can be the one of a galaxy ($z_0\simeq 1.42\times
10^{35}GeV^{-1}$), or cluster of galaxies ($z_0\sim 10^{36}GeV^{-1}$) for example.
Physically meaningful profiles require $1\leq p\leq 2$; $p<1$ implies an
infinite force towards the center of the structure and $p>2$ means that the
density increases towards the outer edges of it. The function $f(z)$ reads 
\begin{equation}
f(z)=\left\{ 
\begin{array}{ccc}
\frac{p\eta _0\alpha }3\frac{z^{p-1}}{z_0^p} & for & z<z_0 \\ 
\frac{p\eta _0\alpha }{3z_0} & for & z>z_0
\end{array}
\right.  \label{ic}
\end{equation}
and 
\begin{equation}
\int^z\frac{dz^{\prime }}{f(z^{\prime })}=\left\{ 
\begin{array}{ccc}
\frac{3z_0^p}{p\eta _0\alpha }\frac 1{(2-p)}z^{2-p} & for & z<z_0 \\ 
\frac{3z_0^2}{p\eta _0\alpha }z & for & z>z_0
\end{array}
\right.  \label{id}
\end{equation}

As stated above, we evaluate the function $H(\zeta )$ by fixing the functional form of 
$\delta n\left( z,\eta \right)  $ at $\eta =\eta _0$. In view of the lack of any 
prescription for it, we also choose it as a constant, i.e. 
$\delta n\left( z,\eta _0\right) = \delta n_0 $
We then have 
\begin{equation}
H\left[ \zeta \right] =\delta n_0\frac{p\eta _0\alpha }{3z_0^p}\left[ \frac{(2-p)p\eta
_0\alpha }{3z_0^p}\zeta \right] ^{\left( p-1\right) /\left( 2-p\right) };\quad
\zeta \leq \frac{3z_0^2}{p\eta _0\alpha }\frac 1{(2-p)}  \label{ie}
\end{equation}

\begin{equation}
H\left[ \zeta \right]=\delta n_0\frac{p\eta _0\alpha }{3z_0};\qquad otherwise  \label{if}
\end{equation}

Finally the density profile $\delta n $ reads

\subparagraph{For $z>z_0$}

\begin{eqnarray}
H\left[ z,\eta \right] &=&\delta n_0\frac{p\eta _0\alpha }{3z_0}  \label{ig} \\
f[z] &=&\frac{p\eta _0\alpha }{3z_0}  \label{ih}
\end{eqnarray}
\begin{equation}
\delta n = \delta n_0  \label{ii}
\end{equation}

\subparagraph{For $z_0^{2-p}-\frac{\alpha p(2-p)\eta _0^2}{3z_0^p}\left( 
\frac{\eta _0}\eta +\frac{\eta ^2}{\eta _0^2}-\frac 32\right)
<z^{2-p}<z_0^{2-p}$}

\begin{eqnarray}
H\left[ z,\eta \right] &=&\delta n_0\frac{\alpha p\eta _0}{3z_0};  \label{ij} \\
f[z] &=&\frac{p\eta _0\alpha }3\frac{z^{p-1}}{z_0^p}  \label{ik}
\end{eqnarray}
\begin{equation}
\delta n = \delta n_0\frac{z_0^{p-1}}{z^{p-1}}  \label{il}
\end{equation}

\subparagraph{For $0<z^{2-p}<z_0^{2-p}-\frac{\alpha p(2-p)\eta _0^2}{3z_0^p}
\left( \frac{\eta _0}\eta +\frac{\eta ^2}{\eta _0^2}-\frac 32\right) $}

\begin{eqnarray}
H\left[ z,\eta \right] &=&\delta n_0\frac{ p\eta _0\alpha }{3z_0}\left[ z^{2-p}+\frac{
(2-p)p\alpha }{3z_0^p}\eta _0^2\left( \frac{\eta _0}\eta +\frac 12\frac{\eta
^2}{\eta _0^2}-\frac 32\right) \right] ^{\left( p-1\right) /\left(
2-p\right) }  \label{im} \\
f[z] &=&\frac{p\eta _0\alpha }3\frac{z^{p-1}}{z_0^p}  \label{in}
\end{eqnarray}
\begin{equation}
\delta n = \delta n_0\frac 1{z^{p-1}}\left[ z^{2-p}+\frac{(2-p)p\alpha }{3z_0^p}\eta
_0^2\left( \frac{\eta _0}\eta +\frac 12\frac{\eta ^2}{\eta _0^2}-\frac
32\right) \right] ^{\left( p-1\right) /\left( 2-p\right) }  \label{io}
\end{equation}

We can see from eq. (\ref{ho}) that the 
magnetic field will grow with time for all exponents in
the physical range $1<p<2$. It remains unperturbed outside the dark matter
distribution, and inside it freezes for long times with a profile given by
eq. (\ref{io}). For $p=1$ we obtain $B=B_0$ for the whole range of $z$. For
example, for $p=3/2$ and the density profile grows like 
\begin{eqnarray}
\delta n &=&\delta n_0\frac 1{\sqrt{z}}\left\{ \sqrt{z}+\frac{\alpha \eta _0^2}{4z_0^{3/2}}
\left[ \frac{\eta _0}\eta +\frac 12\frac{\eta ^2}{\eta _0^2}-\frac 32\right]
\right\}  \label{ip} \\
\ \ for\qquad 0 &<&\sqrt{z}<\sqrt{z_0}-\frac{\alpha \eta _0^2}{4z_0^{3/2}}
\left[ \frac{\eta _0}\eta +\frac 12\frac{\eta ^2}{\eta _0^2}-\frac 32\right]
\nonumber
\end{eqnarray}
\begin{eqnarray}
\delta n &=&\delta n_0\sqrt{\frac{z_0}z}  \label{iq} \\
&&\ for\qquad \sqrt{z_0}-\frac{\alpha \eta _0^2}{4z_0^{3/2}}\left[ \frac{
\eta _0}\eta +\frac 12\frac{\eta ^2}{\eta _0^2}-\frac 32\right] \left. <
\sqrt{z}<\sqrt{z_0}\right.  \nonumber
\end{eqnarray}
and 
\begin{equation}
\delta n = \delta n_0;\qquad for\quad z>z_0  \label{ir}
\end{equation}

We now analyze the case $p=2$.

\paragraph{Homogeneous density contrast.}

Let us consider a dark matter density contrast profile given by 
\begin{equation}
\delta (z) =\left\{ 
\begin{array}{ccc}
\xi & for & z<z_0 \\ 
0 & for & z>z_0
\end{array}
\right.  \label{ja}
\end{equation}

The generalized Newtonian potential can be written as 
\begin{equation}
\psi (z)=\left\{ 
\begin{array}{ccc}
\alpha \left( \frac z{z_0}\right) ^2 & for & z<z_0 \\ 
\alpha \left( 2\frac z{z_0}-1\right) & for & z>z_0
\end{array}
\right.  \label{jb}
\end{equation}

The function $f(z)$ then reads 
\begin{equation}
f(z)=\left\{ 
\begin{array}{ccc}
\frac{2\eta _0\alpha }3\frac z{z_0^2} & for & z<z_0 \\ 
\frac{2\eta _0\alpha }{3z_0} & for & z>z_0
\end{array}
\right.  \label{jc}
\end{equation}
and 
\begin{equation}
\int^z\frac{dz^{\prime }}{f(z^{\prime })}=\left\{ 
\begin{array}{ccc}
\frac{3z_0^2}{2\eta _0\alpha }\ln \left( z\right) & for & z<z_0 \\ 
\frac{3z_0}{2\eta _0\alpha }z & for & z>z_0
\end{array}
\right.  \label{jd}
\end{equation}

With the same initial conditions as before, the function $H\left( \zeta \right) $ 
is given by 
\begin{equation}
H\left[ \zeta \right] =\delta n_0\frac{2\alpha \eta _0B_0}{3z_0^2}\exp \left\{ \frac{2\eta _0\alpha }{
3z_0^2}\zeta \right\} ;\qquad if\quad \zeta \leq \frac{3z_0^2}{2\eta _0\alpha }\ln z_0
\label{je}
\end{equation}
\begin{equation}
H\left[ \zeta \right] =\delta n_0\frac{2\alpha \eta _0}{3z_0};\qquad otherwise  \label{jf}
\end{equation}

The density contrast reads 
\begin{eqnarray}
\delta n &=&\delta n_0\exp \left\{ \frac{2\eta _0^2\alpha }{3z_0^2}\left( \frac{\eta _0}
\eta +\frac 12\frac{\eta ^2}{\eta _0^2}-\frac 32\right) \right\}  \label{jg}
\\
for\qquad 0 &<&z<z_0\exp \left[ -\frac{2\alpha \eta _0^2}{3z_0^2}\left( 
\frac{\eta _0}\eta +\frac 12\frac{\eta ^2}{\eta _0^2}-\frac 32\right) \right]
\nonumber
\end{eqnarray}

\begin{eqnarray}
\delta n &=&\delta n_0\frac{z_0}z;\quad  \label{jh} \\
&&for\quad z_0\exp \left[ -\frac{2\alpha \eta _0^2}{3z_0^2}\left( \frac{\eta
_0}\eta +\frac 12\frac{\eta ^2}{\eta _0^2}-\frac 32\right) \right] \left.
<z<z_0\right.  \nonumber
\end{eqnarray}
and 
\begin{equation}
\delta n = \delta n_0;\qquad for\quad z>z_0  \label{ji}
\end{equation}

\paragraph{Harmonic density profile}

Assume that we have a density distribution whose newtonian potential is
given by

\begin{equation}
\psi (z)=\alpha \left[ \frac 12-\cos \left( 2\pi \frac z{z_0}\right) \right]
\label{ka}
\end{equation}

This function is defined over the whole particle horizon, $z_0$ is a
characteristic scale of the background perturbations, that could be
interpreted as the distance between two adjacent galaxies. The function $
f(z) $ reads 
\begin{equation}
f(z)=\left( \frac{2\pi \alpha \eta _0}{3z_0}\right) \sin \left( \frac{2\pi z
}{z_0}\right)  \label{kb}
\end{equation}
and 
\begin{equation}
\frac{3z_0}{2\pi \alpha \eta _0}\int^z\frac{dz}{\sin \left( 2\pi
z/z_0\right) }=\frac{3z_0^2}{4\pi ^2\alpha \eta _0}\ln \tan \left( \pi \frac
z{z_0}\right)  \label{kc}
\end{equation}

According to what we did before, we have to fix the function 
$H\left[ \zeta \right] $ by
specifying the density contrast at $\eta _0$. We therefore have 
\begin{equation}
\delta n_0=\frac{3z_0}{2\pi \alpha \eta _0\sin \left( \frac{2\pi z}{z_0}\right) }
H\left[ \frac{3z_0^2}{4\pi ^2\alpha \eta _0}\ln \left( \tan \pi \frac
z{z_0}\right) \right]  \label{kd}
\end{equation}
from where we have 
\begin{equation}
H\left[ \zeta \right] =\delta n_0\frac{4\pi \alpha \eta _0}{3z_0}\frac{\exp \left[ \frac{2\pi {\alpha
\eta _0}}{z_0}\zeta \right] }{1+\exp \left[ \frac{4\pi {\alpha \eta _0}}{z_0}
\zeta \right] }  \label{ke}
\end{equation}
And the density profile is
\begin{equation}
\delta n =\delta n_0\frac{\left[ 1+\tan ^2\left( \pi \frac z{z_0}\right) \right]
\exp \left\{ \frac{2\pi {\alpha \eta _0^2}}{3z_0^2}\left( \frac{\eta _0}\eta
+\frac 12\frac{\eta ^2}{\eta _0^2}-\frac 32\right) \right\} }{1+\tan
^2\left( \pi \frac z{z_0}\right) \exp \left\{ \frac{4\pi {\alpha \eta _0^2}}{
3z_0^2}\left( \frac{\eta _0}\eta +\frac 12\frac{\eta ^2}{\eta _0^2}-\frac
32\right) \right\} }  \label{kf}
\end{equation}
We find once again that the density contrast freezes for long times. This time, the
final shape is given by

\begin{equation}
\delta n=\frac {\delta n_0}{\sin ^2\left( \pi \frac z{z_0}\right) }  \label{kg}
\end{equation}
Obviously, the smaller the value of the $\sin $ function, the longer the
time it takes to reach the value given by equation (\ref{kg})

\subsection{Interval of validity of the calculations}

In the case of the power law density profile we see that the magnetic field
diverges for $z\rightarrow 0$ for all times. This is due to the fact that
the density profiles diverges at that point. We will therefore analyze the
limits of validity of the calculations for the case of uniform dark matter density
contrast and of harmonic density.

The first point is to specify until which moment the background
perturbations can be considered in their linear regime. With the value $
\alpha \sim 10^{-5}$ fixed by COBE, we have that the galactic scale entered
its nonlinear regime at a redshift $Z_{nl}^G\sim 5$, which corresponds to $\eta \simeq
15\eta _0$, where $\eta _0\simeq 0.4h^{-1}\times 10^{38}GeV^{-1}$ is the
conformal time at decoupling. For a galaxy cluster, the dark matter is still in
its linear regime (Peebles 1993).

The other assumption we are going to check is neglecting the effect of the
magnetic field on the evolution of the perturbations, namely the the r.h.s
in equation (\ref{ba}). The field affects the motion of the plasma through
the Lorentz force
\begin{equation}
{\cal F}^m={\cal J}_nF^{mn}  \label{la}
\end{equation}
$F^{mn}$ is the e.m. field strength tensor and $J^n$ the induced electric
four current. Since $F^{mn}$ is spacelike to lowest order, the projection
along the four velocity is negligible, and the orthogonal proyection,
written in terms of the conformal magnetic field, yields $\left[ {\bf
J}\times {\bf B}\right] ^i$. The current is given by Maxwell's equations (\ref
{gi}) and (\ref{gj}); after substituting the electric field eq. (\ref{hc}),
we obtain $J^0=0$ (thus no charge separation) and 
\begin{equation}
{\bf J}=\partial _\eta {\bf u}\times {\bf B}+ {\bf u}\times
\partial _\eta {\bf B}+\left( 1-\psi \right) {\bf \nabla }\times {\bf B}-
3{\bf \nabla }\psi \times {\bf B}  \label{lb}
\end{equation}
(it is interesting to note that $\vec J$ is actually orthogonal to the
macroscopic mass flow). For the symmetry of our problem, and neglecting
pressure effects, the equation of motion for matter is changed into

\begin{equation}
\partial _z\psi +\dot u_z+\frac{\dot a}au_z=-\frac{B_y\partial
_zB_y+B_x\partial _zB_x}{a\varepsilon _0}  \label{lc}
\end{equation}
We have to find at what value of $\eta $ the inequality 
\begin{equation}
\left| \partial _z\psi \right| \gg \left| \frac{B_y\partial
_zB_y+B_x\partial _zB_x}{a\varepsilon _0}\right|  \label{ld}
\end{equation}
breaks down. Let us consider the uniform slab. Using equations (\ref{jb})
and (\ref{jh}) we obtain 
\begin{equation}
2\alpha \frac z{z_0^2}\gg \frac{B_0^2z_0^2}{z^3\varepsilon _0}\left( \frac{
\eta _0}\eta \right) ^2\rightarrow \left( \frac \eta {\eta _0}\right) ^2\gg 
\frac{B_0^2z_0^4}{2\alpha z^4\varepsilon _0}  \label{le}
\end{equation}
The correct bound is found by replacing the smallest value of the $z/z_0$
coordinate from equation (\ref{jg}) to obtain 
\begin{equation}
2\alpha \frac z{z_0^2}\gg \frac{B_0^2z_0^2}{z^3\varepsilon _0}\left( \frac{
\eta _0}\eta \right) ^2\rightarrow \left( \frac \eta {\eta _0}\right) ^2\gg 
\frac{B_0^2}{2\alpha \varepsilon _0}\exp \left[ \frac{8\alpha \eta _0^2}{
3z_0^2}\left( \frac{\eta _0}\eta +\frac 12\frac{\eta ^2}{\eta _0^2}-\frac
32\right) \right]  \label{lf}
\end{equation}

As stated before, $\varepsilon _0$ is the critical density of the Universe
at recombination, $\varepsilon _0\sim 8h^2\times 10^{-38}$ GeV$^4$. We take
the value of the primordial field at decoupling as being given by the
mechanisms proposed in the literature for a coherent field at galactic scale
(\cite{dolgov93,ratra92,mazzitelli95,calzetta98,cheng94,sigl97,enqvist94,vachaspati91,dolgov-silk93,hogan83,martin95}), i.e. $B_{today}\sim
10^{-24}Gauss\sim 10^{-45}GeV^2$. We estimate its value at decoupling as $
B_0\sim (1+z_{dec})^2B_{today}\sim 10^{-39}GeV^2$ and we therefore have $
B_0^2/\varepsilon _0\alpha \simeq 10^{-36}$. Replacing in equation (\ref{lf}
) we can check that the inequality is satisfied for $\eta /\eta _0<2.22$ 
($Z\sim 200$) for
a galactic scale $z_0^G\simeq 10^{35}$Gev$^{-1}$ obtaining at that moment
an amplification factor of the order $10^4$. For a cluster scale $
z_0^C\simeq 10^{36}Gev^{-1}$ and equation (\ref{jg}) is valid up to
$\eta /\eta _0 \leq 17.5$ for the same value of $\alpha $,
obtaining a similar amplification factor.

It is also inmediate to check that the neglecting of the last term in 
eq. (\ref{ch}) is a valid approximation for all times.

We therefore conclude that the validity of neglecting the backreation of the
magnetic field on the evolution of the perturbation fixes the time interval 
of validity of our calculations.

\section{Final Remarks}

We have studied the evolution of a preexisting cosmological magnetic field
after recombination of hydrogen, in a background Universe described by a FRW
geometry plus adiabatic perturbations, and considered that the field is
coupled to the remaining ionized fraction. This matter falls into the
background overdensities, dragging with it the magnetic field. This dragging
results in a deviation of the behaviour of the magnetic field from the law
${\cal B} \propto \delta {\cal N}^{2/3}$ where $\delta {\cal N}$ is a 
conserved number density of the plasma,
even when the background perturbations are in their linear regime of 
evolution, as can be seen from equation (\ref{ho}) together with
(\ref{ip}), (\ref{jg}) and (\ref{kf}). The background density
profiles used, although simplistic, retain an important physical
characteristic, namely the asymmetry of the gravitational collapse. In the
case of the harmonic potential, we could also think that it describes an
alternating pattern of walls and voids (\cite{broadhurst90}), despite the 
difference with the scales used in this work. The rate of
growth of the field depends on the characteristics of the background
inhomogeneities and on the initial distribution of magnetic field
intensities. The hypothesis of homogeneity of the field over the horizon
scale at recombination can be sustained provided that the intensity of the
field is sufficiently low, a fact achieved in most of the mechanisms of
primordial field generation (\cite{dolgov93,ratra92,mazzitelli95,calzetta98,cheng94,sigl97,enqvist94,vachaspati91,dolgov-silk93,hogan83,martin95}): 
the larger the coherent scale, the weaker the field intensity. 
The level of amplification depends mainly on the scale of the background
inhomogeneity, as stated in the last subsection, and have its maximum at
the center of the structure. For the $x-y$ plane, the initial symmetry 
remains because gravitational collapse is along the $z$ axis only. As
stated in the previous subsection, our calculations are valid up to
$\eta \simeq 2.22\eta _0$ for galaxies and for the whole time interval
for cluster scales.

Our main result is that it is not correct to consider the magnetic
field evolves as $B\propto B_0/a(\eta )^2$ after recombination. We found
that even in the linear regime of gravitational collapse, there can be
nontrivial corrections to that behaviour. Even if the growth law were a
power law, we could have that the field remains constant between
recombination and the beginning of an amplifying mechanism, over the
scale of interest. Clearly we do not claim to have solved the problem of
the origin of magnetic fields in high redshifted systems, like the damped
Lymann-$\alpha $ clouds, but if the scenario of primordial field generation
plus further amplification is accepted, our results might help to relax 
the constraints on mechanisms for primordial field generation, because 
now the input fields for any further amplifying mechanism could be orders 
of magnitude stronger than the values that those fields would have if 
they had simply decayed as $a^{-2}(\eta )$. For instance, it is stated
(\cite{ruzmaikin90}) that the present intensity of primordial magnetic
field that can seed a galactic dynamo is ${\cal B} \sim 10^{-21}$
Gauss over a scale of 1 Mpc. Most of the proposed mechanism for creation
of the field give a present value 
${\cal B} \sim {\cal O}\left( 10^{-24}\right) $, this means that with the
usual conformal evolution of $B$, the value of the field at recombination
would be ${\cal B}_{rec}\sim 10^{-18}$ Gauss, and at $\eta \sim 2\eta _0$,
${\cal B}\sim 10^{-19}$, while with our calculations, the intensity of the 
field at that moment would be ${\cal B}\sim10^{-16}$ Gauss.

The analysis in this paper may be improved in several particulars, 
the most important ones being the inclusion of vorticity in the plasma
and the extension of our calculations into the nonlinear regime. The others
are assuming a more realistic density profile for the dark matter 
component, and avoiding treating the ionized matter as a fluid by going 
directly to a kinetic description, introducing explicitly the dynamical 
balance between the recombined and ionized fractions of ordinary matter. 
However, as it stands it already shows that there are rich magnetohydrodynamical 
processes occurring as early as $Z\sim 100$, which must be properly understood 
before our picture of the physics of cosmic magnetic fields is complete.

\section{Acknowledgments}

This work has been supported by University of Buenos Aires, CONICET and
Fundaci\'on Antorchas. A. K. wishes to acknowledge Dr. H. Rubinstein and the
hospitality of the University of Stockholm , where part of this work was
completed. E. C. and A. K. thank Oscar Reula for his valuable comments on an
earlier version of this manuscript.

\section{ Appendix}

The Christoffel symbols needed to calculate the covariant derivatives are 
\begin{equation}
\begin{array}{cc}
\Gamma _{00}^0=\frac{\dot a}a+\dot \psi & \Gamma _{0k}^0=\psi ,_k \\ 
\Gamma _{k0}^j=\delta _k^j\left( \frac{\dot a}a-\dot \psi \right) & \Gamma
_{jk}^0=\left[ \frac{\dot a}a\left( 1-4\psi \right) -\dot \psi \right]
\gamma _{jk} \\ 
\Gamma _{jk}^i=\psi ^{/i}\gamma _{jk}-\psi _{/j}\delta _k^i-\psi _{/k}\delta
_j^i & \Gamma _{00}^i=\gamma ^{ik}\psi _{/k}
\end{array}
\label{ma}
\end{equation}
In the context of the paper, these are further simplified, since $\dot \psi
=0$.


\clearpage

\begin{thebibliography}{}
\bibitem[Bardeen et al. 1983]{bardeen83} Bardeen J. M. , Steinhardt P. J. and 
Turner M. S. 1983, \prd 28, 679.

\bibitem[Blasi et al. 1999]{blasi99} Blasi P, Burles S and Olinto A. V. 1999, \apjl 514.

\bibitem[Bond et al 1996]{bond96} Bond J. R., Kofman L. \& Pogosyan D. 1996, \nat  380,
         603.

\bibitem[Brandenberger et al. 1992]{branden92} Brandenberger R. H., 
Feldman H. A. and Mukhanov V. F. 1992, \physrep 215, 203.

\bibitem[Brandenburg  et al. 1997]{brandenburg97} Brandenburg A. , Enqvist K. and
Olesen P. 1997, Phys. Lett. B392, 395.

\bibitem[Broadhurst  et al. 1990]{broadhurst90} Broadhurst T. J. , Ellis R. S., 
Koo D. C. \& Szalay A. S. 1990, Nature 343, 726.

\bibitem[Calzetta et al. 1998]{calzetta98} Calzetta E. A. , Kandus A. and 
Mazzitelli F. D. 1998, \prd 57, 7139.

\bibitem[Cheng \& Olinto 1994]{cheng94} Cheng B. and Olinto A. 1994, \prd 50, 2421.

\bibitem[Dolgov 1993]{dolgov93} Dolgov A. D. 1993, \prd 48, 2499.

\bibitem[Dolgov \& Silk 1993]{dolgov-silk93} Dolgov A. and Silk J. 1993, \prd 47, 3144.

\bibitem[Ellis 1973]{ellis73} Ellis G. F. R. 1973, in  Carg\`ese Lectures in Physics, Vol 6, 
ed. E. Schatzman (Gordon and Breach, New York), 1.

\bibitem[Ellis \& van Elst 1998]{vanelst98} Ellis G.F.R. and van Elst H.1998 ,  Cosmological Models, 
in Carg\`ese Lectures 1998, preprint gr-qc/9812046.

\bibitem[Enqvist \& Olesen 1994]{enqvist94} Enqvist K. and Olesen P. 1994, Phys. Lett. B 329, 195.

\bibitem[Guth \& Pi 1982]{guth82} Guth A. H. and Pi S.-Y. 1982, \prl 49, 1110.

\bibitem[Harrison 1973]{harrison73} Harrison E. R. 1973, \mnras 165, 185.

\bibitem[Hawking 1982]{hawking82} Hawking S. W. 1982, Phys. Lett. B115, 295.

\bibitem[Hogan 1983]{hogan83} Hogan C. 1983, \prl 51, 1488.

\bibitem[Howard \& Kulsrud 1996]{howard96} Howard A. M. and Kulsrud R. M. 1996, preprint astro-ph/9609031.

\bibitem[Israel 1988]{israel88} Israel W. 1988, in  Relativistic fluid dynamics, ed. A. Anile 
        and Y. Choquet - Bruhat (Springer, New York).

\bibitem[Lucchin \& Coles 1995]{luchin} Coles, P. \& Lucchin F. 1995, Cosmology: 
        The Origin and Evolution of Cosmic Structure, (John Wiley \& Sons, New York).

\bibitem[Jedamzik et al. 1998]{jedamzik98} Jedamzik K., Katalinic V. and Olinto A. 1998, 
        \prd 57, 3264.

\bibitem[Martin \& Davies 1995]{martin95} Martin A. P. and Davies A. C. 1995, Phys. Lett. B360, 71.

\bibitem[Mazzitelli \& Spedalieri 1995]{mazzitelli95} Mazzitelli F. D. and Spedalieri F. M. 1995, 
        \prd 52, 6694.

\bibitem[Moffat 1961]{moffat61} Moffatt K. 1961, J. Fluid Mech. 11, 625.

\bibitem[Olesen 1997]{olesen97} Olesen P. 1997, Phys. Lett.  B398, 321.

\bibitem[Oren \& Wolfe 1995]{oren} Oren A. L. and Wolfe A. M. 1995, \apj 445, 624.

\bibitem[Peebles 1993]{peebles93} Peebles P. J. E. 1993, Principles of Physical Cosmology, (NJ, Princeton).

\bibitem[Pudritz \& Silk 1989]{pudritz89} Pudritz R. E. and Silk J. 1989, \apj 342, 650.

\bibitem[Ratra 1992]{ratra92} Ratra B. 1992, \apjl 391, L1.

\bibitem[Ryu et al. 1998]{ryu98} Ryu D., Kang H. and Biermann P. L. 1998, \aap 335

\bibitem[Sigl et al. 1997]{sigl97} Sigl G. , Olinto A. \& Jedamzik K. 1997, 55, 4582.

\bibitem[Starobinskii 1982]{starob82} Starobinskii A. A. 1982, Phys. Lett. B117, 175 (1982).

\bibitem[Subramanian \& Barrow 1998]{subramanian98} Subramanian K. and Barrow J. D. 1998,
        \prd 58, 083502.

\bibitem[Thorne \& MacDonald 1982]{boeing82} Thorne K. S. and MacDonald D. 1982, 
        \mnras  198, Microfiche MN 198/1, 339.

\bibitem[Vachaspati 1991]{vachaspati91} Vachaspati T. 1991, Phys. Lett. B265, 258.

\bibitem[Weinberg 1971]{weinberg71} Weinberg S. 1971, \apj  168, 175.

\bibitem[Wolfe et al. 1992]{wolfe92} Wolfe A. M. , Lanzetta K. and Oren A. L. 1992, \apj 388, 17.

\bibitem[Zel'dovich \& Novikov 1983]{novikov83} Zel'dovich Ya. B. and Novikov I. D. 1983, 
        Relativistic Astrophysics, Vol. 2: The Structure and Evolution of the Universe, 
        (The University of Chicago Press, Illinois).

\bibitem[Zel'dovich et al. 1990]{ruzmaikin90} Zel'dovich Ya. B., Ruzmaikin A. A. 
         and Sokoloff D. D. 1990,  Magnetic Fields in Astrophysics, 2nd. ed., (Gordon and Breach,
         Montreux).
\end{thebibliography}
\end{document}